\documentclass[aps,prd,groupedaddress,reprint, showpacs]{revtex4-1}
\pdfoutput=1
\usepackage{graphicx}
\usepackage[breaklinks, pdftex]{hyperref}
\usepackage[english]{babel}
\usepackage{booktabs}
\graphicspath{{../figures/}}
\usepackage{epstopdf}
\usepackage{rotating}
\usepackage{multirow}
\usepackage{amsmath}
\usepackage{dcolumn}
\usepackage{revsymb}

\begin{document}

\title{Coherent photon scattering background in sub-GeV/$c^2$ direct dark matter searches}

\author{Alan E. Robinson}
\affiliation{Fermi National Accelerator Laboratory, Batavia, Illinois, USA 60510}
\email{fbfree@fnal.gov}

\date{March 17, 2017}
\begin{abstract}
Proposed dark matter detectors with eV-scale sensitivities will detect a large background of atomic (nuclear) recoils from coherent photon scattering of MeV-scale photons.  This background climbs steeply below ${\sim}10$~eV, far exceeding the declining rate of low-energy Compton recoils.  The upcoming generation of dark matter detectors will not be limited by this background, but further development of eV-scale and sub-eV detectors will require strategies, including the use of low nuclear mass target materials, to maximize dark matter sensitivity while minimizing the coherent photon scattering background.
\end{abstract}


\maketitle

Interest in sub-GeV$/c^2$ mass thermal relic dark matter models has inspired ideas for direct detection experiments with eV-scale and sub-eV thresholds \cite{[][{ {a}nd references therein.}]SLAC_workshop}.  For such light dark matter, the recoil energy differential scattering rate is restricted to energies below the detection thresholds of direct detection experiments motivated by weak-scale and supersymmetric models \cite{LUX, models}.


Penetrating MeV-scale photons are a background for all of these experiments with different mechanisms at high and low recoil energies.  Incoherent (Compton) scattering from electrons is suppressed when insufficient energy is deposited to excite a bound electron \cite{Kane199267}, or when the energy range of interest is narrow compared to typical MeV scale energy depositions.  This dominant mechanism for photon backgrounds in most existing direct detection experiments has been considered negligible for future eV-scale and sub-eV experiments \cite{SLAC_workshop}.  In contrast, the coherent scattering of neutral particles, such as coherent neutrino scattering \cite{neutrino} or coherent dark matter scattering, produces an enhanced spectrum of low-energy recoils.  Coherent photon scattering across an atom produces a low-energy background spectrum that may overwhelm low threshold dark matter detectors.


A photon with energy $E_\gamma\,{=}\,c\,p_\gamma \,{\approx}\, 1$~MeV scattering at angle $0 \,{\leq}\, \theta \,{\leq}\, \pi$ from an atom with mass $M\,{\approx}\,10$~GeV/$c^2$, will transfer a small momentum $q$ and recoil energy $E_r$.  
\begin{gather}
\label{eq:1}
E_r = \frac{q^2}{2M} = \frac{(2 p_{\gamma} \sin\frac12 \theta)^2}{2 M} \\
\alt \frac{(2\cdot10^6 \text{ eV}/c)^2} {2 \cdot 10^{10} \text{ eV}/c^2} = 200 \text{ eV} \notag
\end{gather}
The differential coherent scattering cross section is also strongly suppressed when coherence is lost for $q\,{>}\, \hbar/a_B\, {=}\, 3.7$~keV$/c$, where $a_B$ is the Bohr radius.  This small energy deposition can be safely ignored for most applications, but not for upcoming dark matter searches.


\begin{figure}
\centering
\includegraphics[width=\columnwidth]{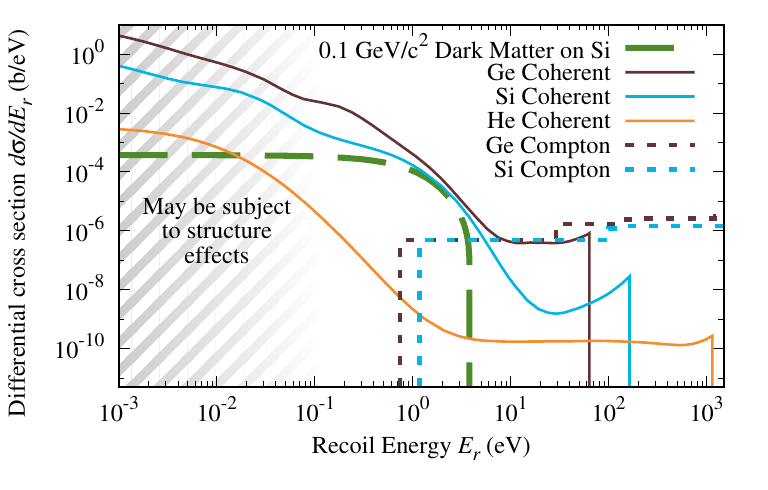}
\caption{\label{fig} Energy differential cross sections for 1461 keV photons from $^{40}$K decay in free silicon, germanium, and helium atoms.  The low energy portion of the coherent scattering spectrum is dominated by Rayleigh scattering, while the high energy components are dominated by nuclear Thomson scattering and Delbr\"uck scattering \cite{Roy19993,KANE198675}.  For low recoil momenta in condensed systems, structure effects may modify the spectrum from the free atom ones shown (see text).  The high energy cutoff of the coherent spectra vary with mass and photon energy as per Equation~\ref{eq:1}.  The Compton scattering spectra shown is an approximation using the Klein-Nishina formula, with cutoffs for the electron shell binding energies and the semiconductor bandgaps \cite{Kittel,CRC,[{For a more thorough treatment of low-energy Compton scattering, see }][]Kane199267,*Monash}.  All energies are given in true recoil energy.  The spectral shape expected in silicon for the elastic scattering of dark matter with 0.1~GeV/c$^2$ mass is shown assuming a Maxwellian dark matter velocity distribution with $v_o\,{=}\,220$~m/s, $v_{esc} \,{=}\, 544$~m/s, and $v_e\,{=}\,244$~m/s \cite{th_di_Lewin_Smith}.}
\end{figure}

\begin{table}
\centering
\begin{ruledtabular}
\begin{tabular}{c c c c}
Recoil Energy Range & \multicolumn{3}{c}{Integrated Scattering Rate} \\
$[$eV$]$ & \multicolumn{3}{c}{$[$recoils (kg$\cdot$yr)$^{-1}]$} \\
\cline{2-4}\\[-1.7ex]
     & Ge & Si & He \\
\hline\\[-1.7ex]
$<$0.01           & 72   & 16    & 1.0 \\
0.01\textendash0.1& 34   & 13    & 0.5 \\
0.1\textendash1   & 16   & 5     & 0.013 \\
1\textendash10    & 0.8  & 0.9   & $1.6\times10^{-4}$ \\
$>$10             & 0.10 & 0.012 & 0.012
\end{tabular}
\end{ruledtabular}
\caption{\label{tab} Expected free atom coherent photon scattering rates in a low-background experiment with a 0.04 counts/kg/day background from low-energy Compton scatter recoils of 1461~keV $^{40}$K decay photons.  This Compton scattering background rate has been demonstrated by the IGEX experiment \cite{IGEX}, and is typical of kg-scale dark matter experiments.  The ratio of the coherent background rate to the low-energy Compton background rate is approximately independent of the photon energy except above the kinetic cutoff energy given by Equation \ref{eq:1}.  At low recoil energies ${\lesssim}\, 0.02$~eV, coherent photon and dark matter scattering rates for atoms in condensed systems will differ from the scattering rates for free atoms (see text).  Below 1~eV, these rates exceed expected rates from coherent neutrino scattering \cite{neutrino}.}
\end{table}

The angle differential cross section for coherent photon scattering and its effects on photon transport have been well studied \cite{Roy19993}\footnote{Coherent photon scattering, neglecting atomic recoils, has been implemented in common radiation transport calculations such as Geant4 \cite{Geant} and MCNP \cite{MCNP}.}.  The dominant Rayleigh scattering cross section can be defined in terms of atomic form factors $F(q,Z)$ multiplying the non-relativistic spin-averaged Thomson scattering cross section~$\sigma_T$ from a single electron with mass $m_e$:
\begin{gather}
\frac{d\sigma_T}{d\Omega} = \frac{e^4}{2 m_e^2 c^4}(1 + \cos^2 \theta) \label{eq:2}\\
\frac{d\sigma}{d\Omega} = \frac{d\sigma_T}{d\Omega} F(q,Z)^2 \label{eq:3}
\end{gather}
where $Z$ is the atomic number and $\Omega$ is the solid angle.  Additional contributions to the scattering amplitude from nuclear Thomson scattering, nuclear resonance scattering, and Delbr\"uck scattering are important at large momentum transfers \cite{[{See }][{ for plots of the scattering contributions from different elastic processes.}]KANE198675}.  By a change of variables, the energy differential cross section is
\begin{equation}
\label{eq}
\frac{d\sigma}{dE_r} = \frac{d\sigma}{d\Omega} \frac{2 \pi M c^2}{E_\gamma^2}
\end{equation}
Detailed scattering-matrix calculations that accurately calculate all contributions to coherent photon scattering are available for free neutral atoms with $Z\geq 13$ and have been validated by many experiments \cite{S-matrix}.  Figure \ref{fig} shows the scattering-matrix calculated cross sections for silicon and germanium.  The cross section for helium is been calculated using non-relativistic form factors for Raleigh scattering \cite{Form_Factor, *Form_Factor_err, EPDL}, Equations \ref{eq:2} and \ref{eq:3} for nuclear Thomson scattering, and lookup tables for Delbr\"uck scattering \cite{Delbruck}. The imaginary Delbr\"uck amplitudes are interpolated in energy by the pair-production cross section \cite{XCOM} while other interpolations from the lookup table are linear.  Expected coherent photon scattering rates for a variety of target materials, assuming a well constructed passive radiation shield \cite{IGEX}, are shown in Table~\ref{tab}.

Within condensed materials, coherence between atoms and phonon quantization modify recoil spectra from those calculated calculated for free atoms shown in Figure \ref{fig} \cite{Kittel, He, *He2}.  Specifically in crystals, coherent scattering may occur across the entire crystal, resulting in no phonon production.  The Bragg scattering intensity $I$ compared to the total scattered intensity $I_o$ is given by the Debye-Waller factor \cite{Kittel}.  This factor may be expressed as,
\begin{equation}
\frac{I}{I_o} = \exp \left (- \frac{ q^2 }{2M}\frac{ \langle U \rangle }{ \langle U_o \rangle^2} \right)
\end{equation}
where $\langle U \rangle$ is the average kinetic energy per lattice site of the crystal, $\langle U_o \rangle \,{\sim} \, 0.02$~eV is the average zero point energy per lattice site \footnote{The average zero point energy can be calculated, assuming a nearly harmonic potential at each lattice site, as half the average phonon energy integrated over the density of phonon states \cite{Kittel}.  For many crystals, the zero point energy is between 0.01 and 0.03 eV \cite{PhysRevB.50.2221, PhysRev.145.492}.  Effects due to phonon quantization are also expressed near this energy scale.}, and $q^2/2M$ is the recoil energy when scattering from a single free atom.  Even at zero temperature in a perfect crystal, where $\langle U \rangle = \langle U_o \rangle$, scattering off the entire crystal is exponentially suppressed for recoil energies greater than the zero point energy.

For scattering processes that produce phonons, a dynamic structure factor, $S(\vec{q},E_r)$, may be defined \cite{He, *He2}.  The double differential scattering rate may be factorize into the contributions from free atom scattering and the structure factor.  While effects described by the structure factor, such as the multi-phonon scattering, may be exploited for dark matter detection, the same factors also modify detector responses to coherent photon scattering.  How these effects are expressed in calorimeters depends on the momentum dependence of the free atom interaction and on the phase space available in the collision \footnote{MeV-scale photons scatter via a point interaction with ${\sim} 3.7$~keV/$c^2$ of momentum available.  At low recoil energies, this interaction is similar to ${\sim}5$~MeV$/c^2$ dark matter interacting via a heavy mediator.}.  

While the cross sections for 1461~keV photons are shown in Figure~\ref{fig}, the ratio of coherent-to-Compton scattering rates can be used to approximate the behaviour of the entire spectrum of background photons in dark matter experiments.  The ratio of Rayleigh, nuclear Thomson, and electron Thomson (non-relativistic Compton) scattering differential cross-sections is fixed within a factor of $1 \leq 1+\cos^2\theta \leq 2$ for any given nuclear target and recoil energy below the kinetic cutoff given in Equation~\ref{eq:1}.

Figure \ref{fig} shows that the differential rate of coherent photon scattering in high-$Z$ materials below recoil energies of 10~eV far exceeds the expected rate from Compton scattering. Upcoming experiments sensitive to GeV$/c^2$-scale dark matter will not be limited by this Raleigh scattering background as their thresholds for nuclear recoils are $\geq 35$eV \cite{CDMS_sensitivity,CRESST}.  Higher energy recoils from nuclear Thomson and Delbr\"uck scattering would be observable if nuclear recoils from coherent photon scattering can be discriminated from electron recoils from Compton scattering and beta decays.  SuperCDMS SNOLAB would need to exceeds its goals for electron recoil / nuclear recoil discrimination to observe these higher energy components of coherent photon scattering.

The energy differential cross-section given by Equation \ref{eq} scale with atomic number, $Z$, and mass, $A$, as $Z^2A$ at $q\,{=}\,0$ and as $Z^4/A$ in the nuclear Thomson scattering regime.  By approximating $A\,{\propto}\, Z$, these coherent photon scattering regimes, spin-independent sub-GeV dark matter-nucleus scattering, and coherent neutrino nuclear scattering all scale identically with target composition \cite{th_di_Lewin_Smith, neutrino}.

At intermediate momenta, coherence for Raleigh scattering depends on how tightly inner shell electrons are bound.  For this recoil energy range, low-$Z$ materials where electrons are weakly bound, such as hydrogen or helium, will have reduced Raleigh scattering backgrounds for a given dark matter sensitivity and flux of MeV-scale photons.  As the Raleigh scattering background climbs steeply for low recoil momenta, the use of lower-$Z$ materials will be as effective as order of magnitude reductions or rejections of the MeV-scale photon flux.

\begin{acknowledgments}
I thank Andreas Biekert for finding the sign error in an earlier version of this work.  This work was completed with the support of the Fermi Research Alliance, LLC under Contract No. De\-AC02\-07CH11359 with the United States Department of Energy. 
\end{acknowledgments}


\bibliography{ref}

\end{document}